\documentclass[journal=jacsat,manuscript=article,layout=twocolumn]{achemso}

\usepackage[version=3]{mhchem} 

\setkeys{acs}{
    doi = true,
}

\usepackage[utf8]{inputenc}
\usepackage[T1]{fontenc}
\usepackage{mathptmx}
\usepackage{etoolbox}
\usepackage{graphicx}
\usepackage{xcolor}
\usepackage{siunitx}
\usepackage{hyperref}
\usepackage{tabularx}
\usepackage{xspace}
\usepackage{comment}

\newcommand{\GO}{\si{\giga\ohm}\xspace}
\newcommand{\MO}{\si{\Mohm}\xspace}
\newcommand{\kO}{\si{\kohm}\xspace}

\newcommand{\R}[1]{\ensuremath{R^\text{LR}_{\text{#1}}}\xspace}
\newcommand{\Vth}{\ensuremath{V_{\text{th}}}\xspace}

\author{Chloé CHOPIN}
    \affiliation{Institute of Condensed Matter and Nanosciences, Universit\'{e} catholique de Louvain, Place Croix du Sud 1, 1348 Louvain-la-Neuve, Belgium}
\author{Simon de WERGIFOSSE}
    \affiliation{Institute of Condensed Matter and Nanosciences, Universit\'{e} catholique de Louvain, Place Croix du Sud 1, 1348 Louvain-la-Neuve, Belgium}
\author{Nicolas MARCHAL}
    \affiliation{Institute of Condensed Matter and Nanosciences, Universit\'{e} catholique de Louvain, Place Croix du Sud 1, 1348 Louvain-la-Neuve, Belgium}
\author{Pascal VAN VELTHEM}
    \affiliation{Institute of Condensed Matter and Nanosciences, Universit\'{e} catholique de Louvain, Place Croix du Sud 1, 1348 Louvain-la-Neuve, Belgium}
\author{Luc PIRAUX}
    \affiliation{Institute of Condensed Matter and Nanosciences, Universit\'{e} catholique de Louvain, Place Croix du Sud 1, 1348 Louvain-la-Neuve, Belgium}
\author{Flavio ABREU ARAUJO}
    \email{flavio.abreuaraujo@uclouvain.be}
    \affiliation{Institute of Condensed Matter and Nanosciences, Universit\'{e} catholique de Louvain, Place Croix du Sud 1, 1348 Louvain-la-Neuve, Belgium}
\title[3D Ag-NW]{Memristive and tunneling effects in 3D interconnected silver nanowires}

\begin{document}
\begin{abstract}
Due to their memristive properties nanowire networks are very promising for neuromorphic computing applications.
Indeed, the resistance of such systems can evolve with the input voltage or current as it confers a synaptic behaviour to the device. 
Here, we propose a network of silver nanowires (Ag-NWs) which are grown in a nanopourous membrane with interconnected nanopores by electrodeposition. 
This bottom-up approach fabrication method gives a conducting network with a 3D architecture and a high density of Ag-NWs. 
The resulting 3D interconnected Ag-NW network exhibits a high initial resistance as well as a memristive behavior. 
It is expected to arise from the creation and the destruction of conducting silver filaments inside the Ag-NW network. 
Moreover, after several cycles of measurement, the resistance of the network switches from a high resistance regime, in the \GO range,  with a tunnel conduction to a low resistance regime, in the \kO range.\\
\end{abstract}


A limitation of current computer architecture is the separation between the memory and the processing unit which leads to the von Neumann bottleneck.
To go beyond this bottleneck, a new field named neuromorphic computing\cite{indiveri2011frontiers} has emerged which takes inspirations from the brain to be more efficient.
Indeed, in the brain, the synapses, which act as memory, and the neurons, which act as processing units, are distributed and the neurons are strongly interconnected through the synapses resulting in a highly parallel architecture.
The complexity of the brain, its high number of interconnections, which can be recurrent, inspire the fabrication of neuromorphic devices\cite{zhu2020comprehensive}.
They are based on various physical principles like spintronics\cite{grollier2020neuromorphic}, optics\cite{shastri2021photonics} or  memristive devices\cite{jo2010nanoscale, avizienis2012neuromorphic, wan2018threshold, diaz2019emergent}.

Memristors were theorised by Chua in 1971 and described as the missing component linking charge and flux\cite{chua1971memristor}.
In 2008, Strukov \textit{et al}\cite{strukov2008missing} showed that some nanodevices exhibiting a non-linear hysteretic $I-V$ curve are in fact memristive devices.
This phenomena was previously observed in memory based on resistive switching \cite{waser2010nanoionics} and memristor were proposed later as neuromorphic synapses\cite{jo2010nanoscale}.
Resistive switching arises from different phenomena\cite{waser2010nanoionics,sawa2008resistive} including the formation and destruction of conductive filaments (CFs) from the migration of metallic ions of Cu\cite{banno2006effect} or Ag\cite{terabe2005quantized} for example.
Atomic switches based on CF are convenient as they can be integrated in a crossbar structure \cite{terabe2005quantized}.
However, in the brain, neurons are highly interconnected by synapses and the connections can be recurrent which leads to a much more complex structure compared to crossbar architecture.
Thus, nanowire networks and especially silver nanowire (Ag-NW) networks\cite{white2011resistive} seem to be good candidates for neuromorphic computing as they have both numerous interconnections and memristive properties.
In addition, Ag-NWs can be used for transparent electronics\cite{hu2010scalable}, have a low electrical resistivity\cite{bellew2015resistance},  and are easily fabricated by different methods\cite{avizienis2012neuromorphic, du2017engineering, milano2019recent}.

\begin{figure*}[!ht]
    \centering
    \includegraphics[scale=1]{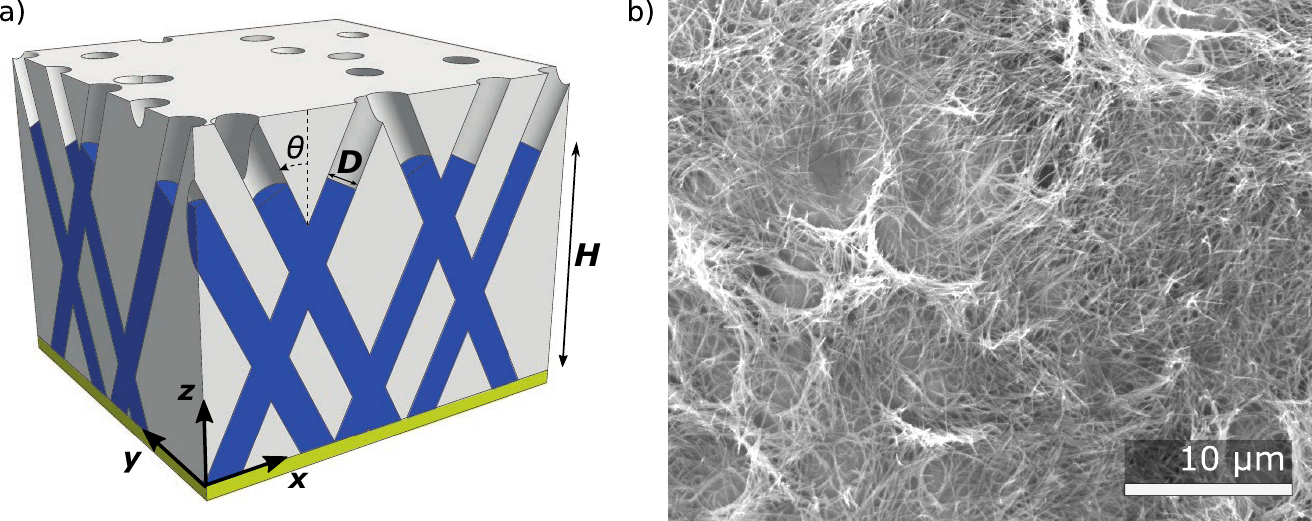}
    \caption{a) Scheme of the 3D interconnected Ag-NW network inside the membrane. The nanowires are represented in blue with $H$ the network height, $D$ the nanowires diameter and $\theta$ the angle with the normal. Some interconnections are shown. b) SEM image of the network after dissolution of the membrane. The network collapses because of the low porosity and the thin diameter of the Ag NWs.}
    \label{fig:network_scheme}
\end{figure*}

Several fabrication processes are used to fabricate Ag-NW network like growing Ag-NWs from Cu seeds\cite{avizienis2012neuromorphic} or with a polyol synthesis\cite{coskun2011polyol} and a polyvinylpyrrolidone (PVP) coating\cite{du2017engineering, wan2018threshold, diaz2019emergent, milano2021materia} (more fabrication processes are detailed in this review\cite{milano2019recent}).
PVP coated Ag-NWs can exhibit different density of Ag-NWs in the sample\cite{diaz2019emergent} as well as a larger resistance because of the PVP coating\cite{du2017engineering}.
Several post-treatments can be used to increase the connections between Ag-NWs including thermal annealing,  mechanical pressing and other techniques\cite{bellew2015resistance, du2017engineering}.
On the contrary, one can use post-treatments to increase the resistance of the network and create opportunities for the creation of Ag CFs so that the network exhibits a memristive behavior.
It can be done either thanks to the encapsulation of the Ag NWs in an insulating shell like PVP\cite{du2017engineering, diaz2019emergent, milano2021materia} where conductive filaments grow across the PVP insulating layer\cite{milano2021materia} or by fragmenting the Ag-NWs with an UV/ozone irradiation followed by an annealing\cite{wan2018threshold}.

In this work, we propose a memristive device made of three-dimensional (3D) interconnected Ag-NWs.
Those are deposited inside a nanoporous membrane which allows a bottom-up fabrication of a random but ordered NW network with a high number of interconnections\cite{rauber2011highly}.

\begin{figure}[!ht]
    \centering
    \includegraphics[scale=1]{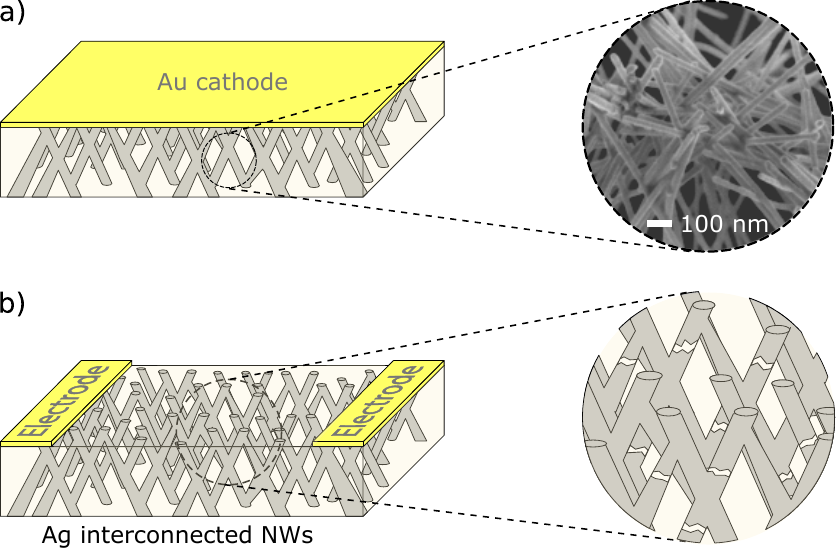}
    \caption{Scheme of the 3D interconnected Ag-NW network and the Au cathode. (a) Before the Au cathode etching, the NW network is intact. Inset: SEM image of the Ag NW network after dissolution of the membrane. (b) After the Au cathode etching, only two electrodes remain. Inset: the etching process has damaged the nanowires. The damages are represented as gaps. The damaged network is not imaged by SEM as it would collapse once the membrane is removed.}
    \label{fig:scheme}
\end{figure}

An electrochemical deposition in ion track nanoporous polycarbonate (PC) membrane was employed to produce 3D interconnected Ag-NWs as shown schematically in fig.\ref{fig:network_scheme}(a).
The PC nanoporous membrane with interconnected nanopores has been fabricated by exposing a \SI{25}{\micro\meter}-thick PC film to a two-step irradiation process. 
The topology of the membrane was defined by exposing the film to a first irradiation step at two fixed angles of \ang{-25} and \ang{+25} with respect to the normal axis of the film plane.
In pratice, the angles are between \ang{20} and \ang{25}.
After rotating the PC film in the plane by \ang{90}, the second irradiation step took place at the same fixed angular irradiation flux to finally form a 3D nanoporous network. 
The diameter of the latent tracks was enlarged by following a previously reported protocol\cite{ferain2003track} to obtain a membrane with an average pore diameter of  30~nm and a volumetric porosity of about 0.4\%. 
Next, the PC templates were coated on one side using an e-beam evaporator with a metallic Cr~(3~nm)/Au~(250~nm) bilayer to serve as a cathode during the electrochemical deposition.
The Ag-NWs were fabricated by electrodeposition using a silver-cyanide-based commercial electrolyte (Silver-Bright-100, Metaken GmbH) in a two-electrode configuration at room temperature by applying a constant potential of -1.5~V versus a double-junction Ag/AgCl reference electrode and a platinum strip used as a counter electrode. 
The reaction is the following:
\begin{equation}
    \text{Ag(SCN)}^{2-}_{3(\text{aq})}+\text{e}^- \rightleftharpoons \text{Ag}_{(\text{s})} + 3\text{SCN}^{-}_{(\text{aq})}
\end{equation}
After the silver electrodeposition process, the network height $H$ is approximately \SI{24}{\micro \meter}.
A schematic view of the sample is presented in fig.\ref{fig:network_scheme}(a).
The morphology of the interconnected Ag-NW network was characterized using a field-emission scanning electron microscope (FE-SEM). For the electron microscopy analysis, the PC template was removed by chemical dissolution using dichloromethane from Sigma-Aldrich. 
As the network porosity is low, the network collapses once the membrane is completely removed as shown on the SEM image in fig.~\ref{fig:network_scheme}(b).
A scheme of the Ag-NWs encapsulated in the membrane with the Au cathode is presented in fig.~\ref{fig:scheme}(a) while its inset shows a closer and tilted view of SEM image of the 3D  interconnected Ag-NW network.
As it can be seen, the Ag-NW network has a complex interconnected structure.

The 3D structure presents a high number of interconnections with the possibility of recurrent connections with the advantages of NWs with a regular diameter and length.
The number of interconnections is estimated numerically and gives 5.5$\cdot 10^8$ $\pm$ 1.0$\cdot 10^8$  interconnections per \si{\milli\meter\cubed}.
For the network height, the density is around 10$^7$ interconnections per \si{\milli\meter\squared} which is one order of magnitude greater than this recent structure\cite{milano2020brain}.
The number of interconnection is easily adapted by adjusting the NWs diameter as well as the membrane porosity.
It is interesting to note that the interconnections between Ag NWs share different portion of volume from none (\textit{i.e.} no crossing) to a complete crossing.

Previously reported 3D interconnected NW network\cite{da2019tunable} with NWs made of Ni with a diameter of 80~nm and a volumetric porosity of 3\% have reached resistance as low as \SI{4}{\ohm}.
To increase the resistance of the 3D interconnected Ag-NW network and see a memristive behavior, the Ag-NWs needs to be damaged.
This is achieved during the etching process where   the cathode is partially removed to create a two electrodes device for electrical measurements\cite{gomes2019making} while insulating domains are created in the network (see fig.~\ref{fig:scheme}(b)).
Indeed, the plasma etching also led to some heating of the polymer membrane which is expected to generate gaps and holes in the sparse network of Ag-NW resulting in a sample with a high initial resistance.
The resistance of the 3D interconnected Ag-NW network is then in the \GO range and above.
The etching process is thus considered as a functionalizing step.

This fabrication process allows to produce NWs below the lithography limit with numerous interconnections and a complexity beyond crossbar architecture as well as providing an easy  scale-up of the system by simply increasing its surface.
As the Ag-NWs are already interconnected, the membrane is not dissolved at the end of the fabrication.
Thus, the network has an enhanced solidity, is flexible\cite{da2021magneto} and is protected against oxidation.
In addition, the randomness needed for neuromorphic application is conserved due to the random location of nanopores in the membrane, the variety of crossing types and the functionalizing step.
This method is low-cost, reliable and both diameter and  density of Ag-NWs can be easily tuned.

\begin{figure}[!ht]
    \centering
    \includegraphics[scale=1]{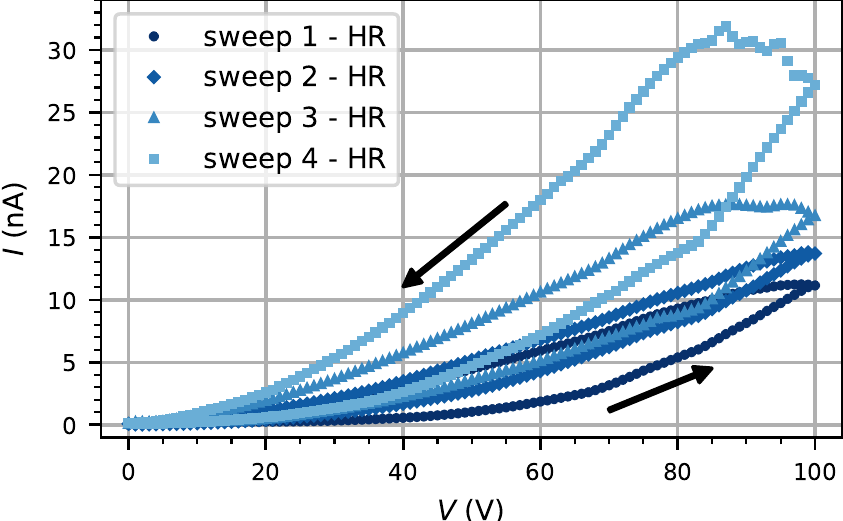}
    \caption{Consecutive $I-V$ cycles in the high resistance regime. A triangular voltage ramp is used. It starts at 0~V, increases to 100~V and finally decreases back to 0~V. The arrows show the evolution of the hysteresis with the voltage. A pinched hysteresis, typical of a memristive behavior, can be seen for each consecutive sweep.
    }
    \label{fig:HR_IV}
\end{figure}

The measured sample is about 5~mm long and 1~mm wide, and the electrical contacts were directly made by silver paint. The $I-V$ curves of the Ag-NWs were obtained using a Keithley 617 electrometer.
All measurements are made at room temperature and tests were made to discard any electrical conduction \textit{via} the PC membrane to ensure a complete conduction through the Ag-NWs and their interconnections only.
First, an $I-V$ cycle is repeated several times with a voltage ramp increasing from 0~V to 100~V before decreasing back to \SI{0}{\volt} with a time step of 1~V/s (see Fig.~\ref{fig:HR_IV}).
The data are denoised thanks to a digital wavelet transform\cite{Lee2019} and the same offset is added to each sweep to ensure a positive current.
Four consecutive cycles are plotted in Fig.~\ref{fig:HR_IV}).
One can see the typical behavior of a memristive system with a pinched hysteresis for all measurements. 
Similar results were obtain in bulk silver nanowire–polystyrene
composites\cite{white2011resistive} with an applied voltage ramp up to 160~V and with PVP-coated Ag-NW network \cite{diaz2020associative} for a threshold voltage below 2V.
We expect that the memristive behaviour arises from the creation of field-induced Ag CF in the holes and gaps created during the plasma etching when the voltage increases.
However, when the voltage decreases, the new filaments break due to Rayleigh instability\cite{wang2019surface} and, hence, the resistance increases.
Indeed, the atoms in the filaments are less stable than the atoms in bulk nanowires, thus the creation of silver filaments is volatile.

For each cycle, there is a local minimum resistance.
Their values are 8.40 $\pm$ 0.02~\GO,  7.11 $\pm$ 0.04~\GO,  4.84 $\pm$ 0.02~\GO and  2.73 $\pm$ 0.02~\GO, respectively.
One can notice that the local minimum resistance decreases with each consecutive cycle.
Even with a voltage ramp reaching a high voltage (100~V), as the current flowing in the device is in the nA range, the maximal power reach for each sweep is respectively \SI{1.12}{\micro\watt}, \SI{1.37}{\micro\watt}, \SI{1.7}{\micro\watt} and \SI{2.9}{\micro\watt}.
This device is then resilient to high input voltage with a moderate power consumption.

\begin{figure}[!ht]
    \centering
    \includegraphics[scale=1]{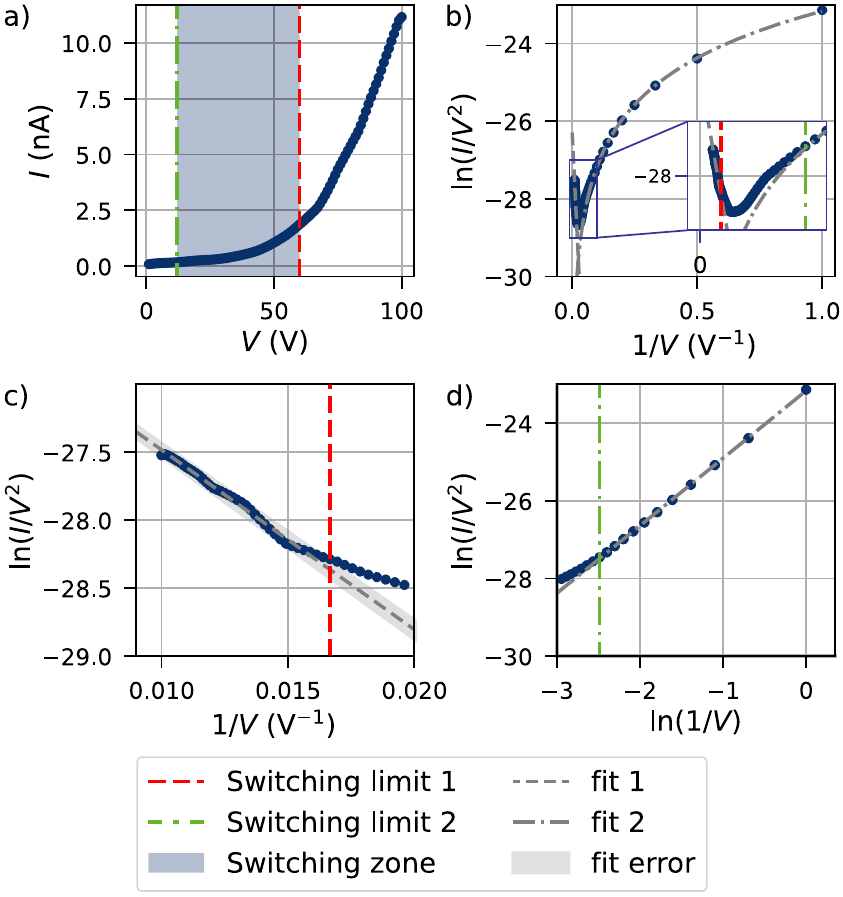}
    \caption{Tunneling phenomena in sweep 1 - HR. (a) Data are extracted from sweep 1 until $V$ = \SI{100}{\volt}. (b) Fowler-Nordheim plot from data presented in (a). Inset: zoom on the switching region. (c) Field-emission at high voltage (d) Direct tunneling at low voltage. Fits 1 and 2 are linear fitting with $y = ax +b$ and the fitting error limits are determine as $y = (a\pm\Delta a)x + (b\pm\Delta b)$ with $\Delta a$ and $\Delta b$ the errors on the parameters $a$ and $b$ after fitting. }
    \label{fig:tunneling}
\end{figure}

For the first sweep, a change of the conduction occurs when a threshold voltage \Vth is exceeded which leads to a strong current increase. Below this threshold, the increase of the current $I$ with the voltage $V$ is much smaller.
This can be explained by the switching between direct tunneling at low voltage and field emission at high voltage as a Fowler-Nordheim plot suggests \cite{araidai2010theoretical} (see Fig.~\ref{fig:tunneling}(b)).
Similar results have been reported with another type of Ag-NW network\cite{wan2018threshold}.
We suppose that the tunnel conduction arises when nanowires are separated by an empty zone or gap acting like an insulating barrier which appears during the etching of the cathode.

In order to have an estimation of the switching zone between these two regimes, a linear fitting is performed on a selection of data for both field emission and direct tunneling (respectively "fit 1" and "fit 2" on Fig.~\ref{fig:tunneling}). 
For the field emission, a linear dependence is shown when $\text{ln}(I/V^2)$ is plotted as a function of $1/V$ as it appears in Fig.~\ref{fig:tunneling}(c).
The direct tunneling exhibits a linear dependence when $\text{ln}(I/V^2)$ is plotted as a function of $\text{ln}(1/V)$ as shown in Fig.~\ref{fig:tunneling}(d).
An error is extracted from "fit 1" (resp. "fit 2") and the switching limits are determined as the first (resp. last) value outside the fitting error range. The error of "fit 2" is small and is hardly seen in Fig.~\ref{fig:tunneling}(d). 
These two switching limits are  12 and 60~V for the direct tunneling and field emission respectively.
The switching of the tunneling conduction arises from a change of the tunneling barrier shape from a rectangular barrier to a triangular barrier\cite{araidai2010theoretical} which  enhances the probability of tunneling.

We repeatedly measured  this sample with $I-V$ cycles, until its resistance dropped in the \MO range and below. This new regime is called low resistance regime (LR) due to the decrease of the minimum resistance by a factor of about 47,000.
After several repeated $I-V$ cycles, we suppose that some filaments present an extended lifetime leading to a strong decrease of the resistance.
A similar behaviour was reported by Avizienis \textit{et al}\cite{avizienis2012neuromorphic}.
\\

\begin{figure}[!ht]
    \centering
    \includegraphics[scale=1]{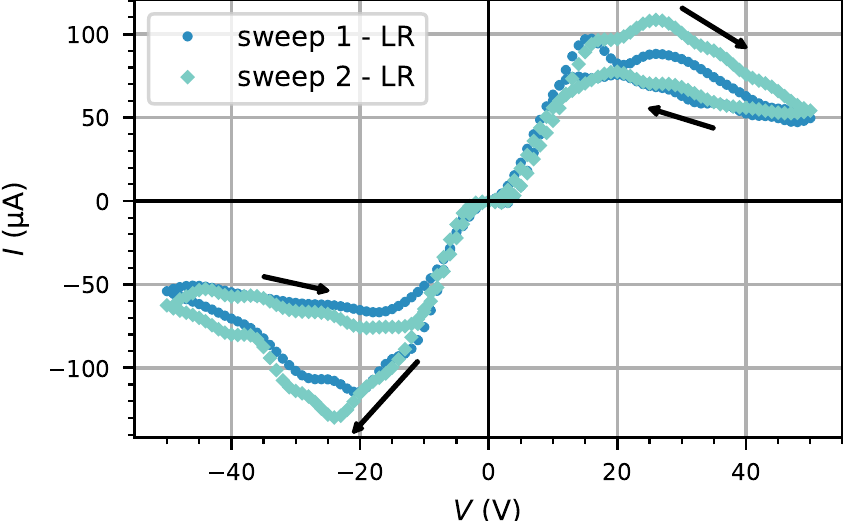}
    \caption{$I-V$ cycles for two consecutive measurements in the low resistance regime. The voltage ramp starts at 0~V, increases to 50~V, then decreases to -50~V and finally increases back to 0~V. The arrows show the direction of the current evolution with the voltage ramp.
    }
    \label{fig:LR_IV}
\end{figure}

Several measurements are made in the low resistance regime and two consecutive $I-V$ cycles are reported in Fig.~\ref{fig:LR_IV}.
The data are denoised in two steps.
First, outliers above \SI{200}{\micro\ampere} are suppressed and replaced  thanks to an interpolation. 
Then, the data are denoised with a digital wavelet transform\cite{Lee2019}.
As seen in Fig.~\ref{fig:LR_IV}, a different memristive behavior is observed compared to the high resistance regime.

The voltage ramp starts at 0~V, increases to 50~V before decreasing to -50~V and returning to 0~V.
The $I-V$ curves show several interesting behaviors.
At first, the current increases with the voltage for $V>2$ V until $V \sim 10-25$ V depending on the sweep.
Then, the current decreases while the voltage continues to increase until $V = 50$ V is reached.
Then, the current increases while the voltage decreases until $V \sim$ 15-20~V and finally, the current decreases with the voltage until $V = 0$~V. The same repeats in a symmetric way when the voltage goes down to -50 V and back to 0 V.

This induces that during the first half of the sweep, a local maximum and two local minima of resistance are reached.
The corresponding resistances are given in Table~\ref{tab:resistance} before being averaged.
The resulting averaged minimum resistance is $\R{min}= 0.163 \pm 0.021$~\MO while the following averaged maximum resistance is obtained: $\R{max}= 0.965 \pm 0.083$~\MO. 
There is a factor six between the minimum and the maximum averaged resistance in the first-half of the sweeps.
Similar results were obtained in the second half of the sweeps with $\R{min}= 0.156\pm0.046$~\MO and $\R{max}= 0.862\pm0.123$~\MO.
In this low resistance regime, the minimum resistances are in the \kO range, with a voltage threshold in the order of 2 V.
\\

\begin{table}[!ht]
    \caption{Minimum and maximum resistances in the low resistance regime and their associate voltage during the first half of the sweeps. \R{min,fwd} is measured in the first part of the first sweep when $V$ increases from 0~V to 50~V, while \R{min,bwd} is measured in the second part when $V$ decreases from 50~V to 0~V. The maximal resistances are taken at 50~V for clarity. 
    }
    \label{tab:resistance}
    \begin{tabular}{lcc}
        \hline
          & Sweep 1 - LR & Sweep 2 - LR \\ \hline
        \R{min, fwd} & 0.150~\MO at 13~V & 0.168~\MO at 15~V\\ 
         \R{max}  & 1.007~\MO at 50~V & 0.923~\MO at 50~V\\ 
         \R{min, bwd} & 0.157~\MO at 10~V & 0.178~\MO at 9~V\\
    \end{tabular}
\end{table}


In conclusion, we reported a 3D interconnected Ag-NW network with a simple two step fabrication process composed of a template-assisted electrodeposition and an etching leading to a highly interconnected network.
The latter has a double function as it both damages the Ag NWs creating holes and gaps were Ag CF can grow, giving rise to memristive properties to the 3D interconnected Ag NW network, and defines electrode pads for electrical connection.
In the future, the etching pattern will be tuned from a basic two electrodes design to a more elaborate pattern with numerous input/output pads.
The resulting NWs are highly ordered and encapsulated in a 3D  nanoporous polymer film.
The 3D architecture allows to have 5.5 $\pm$ 1.0 10$^8$ interconnections per \si{\milli\meter\cubed} with an increased complexity as the interconnections are random with a variety of crossing type.
Two resistance regimes are measured with different resistance range: in the \GO range and above for the high resistance regime and in the \MO range and below for the low resistance regime.
A tunneling conduction arises in the high resistance regime and both high and low resistance regimes exhibit a memristive behavior.
The presented 3D interconnected Ag-NW network could be used for neuromorphic applications due to its memristive properties and more precisely for reservoir computing in a similar way as these recent studies\cite{kuncic2020neuromorphic, milano2021materia}.
In addition, the high density of interconnections will be exploited to train multiple learning process on the same device as numerous pathway are expected to growth in the Ag-NWs network\cite{diaz2020associative}.
Thus, it is a promising device for future application in neuromorphic computing.

\section*{Author's contribution}
F.A.A. and L.P. designed the study.
N.M., L.P., and P.V.V. fabricated the samples. 
F.A.A. conceived and designed the experimental measurement setup and the analysis.
F.A.A., C.C. and S.d.W. collected the data and performed the analysis. 
C.C. wrote the core of the manuscript and all the other co-authors contributed to it.

\begin{acknowledgement}
F.A.A. is a Research Associate of the F.R.S.-FNRS. 
S.d.W. aknowledges the Walloon Region and UCLouvain for FSR financial support.
N.M. acknowledges the Research Science Foundation of Belgium (F.R.S.-FNRS) for financial support (FRIA grant).
The authors would like to thank Dr. E. Ferain and the it4ip Company for supplying polycarbonate membranes.

\end{acknowledgement}


\newcommand{\doi}[1]{\href{http://dx.doi.org/#1}{\nolinkurl{#1}}}


\begin{mcitethebibliography}{33}
    \providecommand*\natexlab[1]{#1}
    \providecommand*\mciteSetBstSublistMode[1]{}
    \providecommand*\mciteSetBstMaxWidthForm[2]{}
    \providecommand*\mciteBstWouldAddEndPuncttrue
    {\def\EndOfBibitem{\unskip.}}
    \providecommand*\mciteBstWouldAddEndPunctfalse
    {\let\EndOfBibitem\relax}
    \providecommand*\mciteSetBstMidEndSepPunct[3]{}
    \providecommand*\mciteSetBstSublistLabelBeginEnd[3]{}
    \providecommand*\EndOfBibitem{}
    \mciteSetBstSublistMode{f}
    \mciteSetBstMaxWidthForm{subitem}{(\alph{mcitesubitemcount})}
    \mciteSetBstSublistLabelBeginEnd
    {\mcitemaxwidthsubitemform\space}
    {\relax}
    {\relax}
    
    \bibitem[Indiveri and Horiuchi(2011)Indiveri, and
    Horiuchi]{indiveri2011frontiers}
    Indiveri,~G.; Horiuchi,~T.~K. Frontiers in neuromorphic engineering.
    \emph{Frontiers in neuroscience} \textbf{2011}, \emph{5}, 118, DOI:
    \doi{10.3389/fnins.2011.00118}\relax
    \mciteBstWouldAddEndPuncttrue
    \mciteSetBstMidEndSepPunct{\mcitedefaultmidpunct}
    {\mcitedefaultendpunct}{\mcitedefaultseppunct}\relax
    \EndOfBibitem
    \bibitem[Zhu \latin{et~al.}(2020)Zhu, Zhang, Yang, and
    Huang]{zhu2020comprehensive}
    Zhu,~J.; Zhang,~T.; Yang,~Y.; Huang,~R. A comprehensive review on emerging
    artificial neuromorphic devices. \emph{Applied Physics Reviews}
    \textbf{2020}, \emph{7}, 011312, DOI: \doi{10.1063/1.5118217}\relax
    \mciteBstWouldAddEndPuncttrue
    \mciteSetBstMidEndSepPunct{\mcitedefaultmidpunct}
    {\mcitedefaultendpunct}{\mcitedefaultseppunct}\relax
    \EndOfBibitem
    \bibitem[Grollier \latin{et~al.}(2020)Grollier, Querlioz, Camsari,
    Everschor-Sitte, Fukami, and Stiles]{grollier2020neuromorphic}
    Grollier,~J.; Querlioz,~D.; Camsari,~K.; Everschor-Sitte,~K.; Fukami,~S.;
    Stiles,~M.~D. Neuromorphic spintronics. \emph{Nature electronics}
    \textbf{2020}, \emph{3}, 360--370, DOI: \doi{10.1038/s41928-019-0360-9}\relax
    \mciteBstWouldAddEndPuncttrue
    \mciteSetBstMidEndSepPunct{\mcitedefaultmidpunct}
    {\mcitedefaultendpunct}{\mcitedefaultseppunct}\relax
    \EndOfBibitem
    \bibitem[Shastri \latin{et~al.}(2021)Shastri, Tait, de~Lima, Pernice,
    Bhaskaran, Wright, and Prucnal]{shastri2021photonics}
    Shastri,~B.~J.; Tait,~A.~N.; de~Lima,~T.~F.; Pernice,~W.~H.; Bhaskaran,~H.;
    Wright,~C.~D.; Prucnal,~P.~R. Photonics for artificial intelligence and
    neuromorphic computing. \emph{Nature Photonics} \textbf{2021}, \emph{15},
    102--114, DOI: \doi{10.1038/s41566-020-00754-y}\relax
    \mciteBstWouldAddEndPuncttrue
    \mciteSetBstMidEndSepPunct{\mcitedefaultmidpunct}
    {\mcitedefaultendpunct}{\mcitedefaultseppunct}\relax
    \EndOfBibitem
    \bibitem[Jo \latin{et~al.}(2010)Jo, Chang, Ebong, Bhadviya, Mazumder, and
    Lu]{jo2010nanoscale}
    Jo,~S.~H.; Chang,~T.; Ebong,~I.; Bhadviya,~B.~B.; Mazumder,~P.; Lu,~W.
    Nanoscale memristor device as synapse in neuromorphic systems. \emph{Nano
        letters} \textbf{2010}, \emph{10}, 1297--1301, DOI:
    \doi{10.1021/nl904092h}\relax
    \mciteBstWouldAddEndPuncttrue
    \mciteSetBstMidEndSepPunct{\mcitedefaultmidpunct}
    {\mcitedefaultendpunct}{\mcitedefaultseppunct}\relax
    \EndOfBibitem
    \bibitem[Avizienis \latin{et~al.}(2012)Avizienis, Sillin, Martin-Olmos, Shieh,
    Aono, Stieg, and Gimzewski]{avizienis2012neuromorphic}
    Avizienis,~A.~V.; Sillin,~H.~O.; Martin-Olmos,~C.; Shieh,~H.~H.; Aono,~M.;
    Stieg,~A.~Z.; Gimzewski,~J.~K. Neuromorphic atomic switch networks.
    \emph{PLoS ONE} \textbf{2012}, DOI: \doi{10.1371/journal.pone.0042772}\relax
    \mciteBstWouldAddEndPuncttrue
    \mciteSetBstMidEndSepPunct{\mcitedefaultmidpunct}
    {\mcitedefaultendpunct}{\mcitedefaultseppunct}\relax
    \EndOfBibitem
    \bibitem[Wan \latin{et~al.}(2018)Wan, Pan, Du, Qu, Yi, and
    Chu]{wan2018threshold}
    Wan,~T.; Pan,~Y.; Du,~H.; Qu,~B.; Yi,~J.; Chu,~D. Threshold switching induced
    by controllable fragmentation in silver nanowire networks. \emph{ACS applied
        materials \& interfaces} \textbf{2018}, \emph{10}, 2716--2724, DOI:
    \doi{10.1021/acsami.7b16142}\relax
    \mciteBstWouldAddEndPuncttrue
    \mciteSetBstMidEndSepPunct{\mcitedefaultmidpunct}
    {\mcitedefaultendpunct}{\mcitedefaultseppunct}\relax
    \EndOfBibitem
    \bibitem[Diaz-Alvarez \latin{et~al.}(2019)Diaz-Alvarez, Higuchi, Sanz-Leon,
    Marcus, Shingaya, Stieg, Gimzewski, Kuncic, and Nakayama]{diaz2019emergent}
    Diaz-Alvarez,~A.; Higuchi,~R.; Sanz-Leon,~P.; Marcus,~I.; Shingaya,~Y.;
    Stieg,~A.~Z.; Gimzewski,~J.~K.; Kuncic,~Z.; Nakayama,~T. Emergent dynamics of
    neuromorphic nanowire networks. \emph{Scientific reports} \textbf{2019},
    \emph{9}, 1--13, DOI: \doi{10.1038/s41598-019-51330-6}\relax
    \mciteBstWouldAddEndPuncttrue
    \mciteSetBstMidEndSepPunct{\mcitedefaultmidpunct}
    {\mcitedefaultendpunct}{\mcitedefaultseppunct}\relax
    \EndOfBibitem
    \bibitem[Chua(1971)]{chua1971memristor}
    Chua,~L. Memristor-the missing circuit element. \emph{IEEE Transactions on
        circuit theory} \textbf{1971}, \emph{18}, 507--519, DOI:
    \doi{10.1109/TCT.1971.1083337}\relax
    \mciteBstWouldAddEndPuncttrue
    \mciteSetBstMidEndSepPunct{\mcitedefaultmidpunct}
    {\mcitedefaultendpunct}{\mcitedefaultseppunct}\relax
    \EndOfBibitem
    \bibitem[Strukov \latin{et~al.}(2008)Strukov, Snider, Stewart, and
    Williams]{strukov2008missing}
    Strukov,~D.~B.; Snider,~G.~S.; Stewart,~D.~R.; Williams,~R.~S. The missing
    memristor found. \emph{nature} \textbf{2008}, \emph{453}, 80--83, DOI:
    \doi{10.1038/nature06932}\relax
    \mciteBstWouldAddEndPuncttrue
    \mciteSetBstMidEndSepPunct{\mcitedefaultmidpunct}
    {\mcitedefaultendpunct}{\mcitedefaultseppunct}\relax
    \EndOfBibitem
    \bibitem[Waser and Aono(2010)Waser, and Aono]{waser2010nanoionics}
    Waser,~R.; Aono,~M. \emph{Nanoscience And Technology: A Collection of Reviews
        from Nature Journals}; World Scientific, 2010; pp 158--165, DOI:
    \doi{10.1142/9789814287005_0016}\relax
    \mciteBstWouldAddEndPuncttrue
    \mciteSetBstMidEndSepPunct{\mcitedefaultmidpunct}
    {\mcitedefaultendpunct}{\mcitedefaultseppunct}\relax
    \EndOfBibitem
    \bibitem[Sawa(2008)]{sawa2008resistive}
    Sawa,~A. Resistive switching in transition metal oxides. \emph{Materials today}
    \textbf{2008}, \emph{11}, 28--36, DOI:
    \doi{10.1016/S1369-7021(08)70119-6}\relax
    \mciteBstWouldAddEndPuncttrue
    \mciteSetBstMidEndSepPunct{\mcitedefaultmidpunct}
    {\mcitedefaultendpunct}{\mcitedefaultseppunct}\relax
    \EndOfBibitem
    \bibitem[Banno \latin{et~al.}(2006)Banno, Sakamoto, Hasegawa, Terabe, and
    Aono]{banno2006effect}
    Banno,~N.; Sakamoto,~T.; Hasegawa,~T.; Terabe,~K.; Aono,~M. Effect of ion
    diffusion on switching voltage of solid-electrolyte nanometer switch.
    \emph{Japanese journal of applied physics} \textbf{2006}, \emph{45}, 3666,
    DOI: \doi{10.1143/jjap.45.3666}\relax
    \mciteBstWouldAddEndPuncttrue
    \mciteSetBstMidEndSepPunct{\mcitedefaultmidpunct}
    {\mcitedefaultendpunct}{\mcitedefaultseppunct}\relax
    \EndOfBibitem
    \bibitem[Terabe \latin{et~al.}(2005)Terabe, Hasegawa, Nakayama, and
    Aono]{terabe2005quantized}
    Terabe,~K.; Hasegawa,~T.; Nakayama,~T.; Aono,~M. Quantized conductance atomic
    switch. \emph{Nature} \textbf{2005}, \emph{433}, 47--50, DOI:
    \doi{10.1038/nature03190}\relax
    \mciteBstWouldAddEndPuncttrue
    \mciteSetBstMidEndSepPunct{\mcitedefaultmidpunct}
    {\mcitedefaultendpunct}{\mcitedefaultseppunct}\relax
    \EndOfBibitem
    \bibitem[White \latin{et~al.}(2011)White, Vora, Kikkawa, and
    Winey]{white2011resistive}
    White,~S.~I.; Vora,~P.~M.; Kikkawa,~J.~M.; Winey,~K.~I. Resistive switching in
    bulk silver nanowire--polystyrene composites. \emph{Advanced Functional
        Materials} \textbf{2011}, \emph{21}, 233--240, DOI:
    \doi{10.1002/adfm.201001383}\relax
    \mciteBstWouldAddEndPuncttrue
    \mciteSetBstMidEndSepPunct{\mcitedefaultmidpunct}
    {\mcitedefaultendpunct}{\mcitedefaultseppunct}\relax
    \EndOfBibitem
    \bibitem[Hu \latin{et~al.}(2010)Hu, Kim, Lee, Peumans, and Cui]{hu2010scalable}
    Hu,~L.; Kim,~H.~S.; Lee,~J.-Y.; Peumans,~P.; Cui,~Y. Scalable coating and
    properties of transparent, flexible, silver nanowire electrodes. \emph{ACS
        nano} \textbf{2010}, \emph{4}, 2955--2963, DOI: \doi{10.1021/nn1005232}\relax
    \mciteBstWouldAddEndPuncttrue
    \mciteSetBstMidEndSepPunct{\mcitedefaultmidpunct}
    {\mcitedefaultendpunct}{\mcitedefaultseppunct}\relax
    \EndOfBibitem
    \bibitem[Bellew \latin{et~al.}(2015)Bellew, Manning, Gomes~da Rocha, Ferreira,
    and Boland]{bellew2015resistance}
    Bellew,~A.~T.; Manning,~H.~G.; Gomes~da Rocha,~C.; Ferreira,~M.~S.;
    Boland,~J.~J. Resistance of single Ag nanowire junctions and their role in
    the conductivity of nanowire networks. \emph{ACS nano} \textbf{2015},
    \emph{9}, 11422--11429, DOI: \doi{10.1021/acsnano.5b05469}\relax
    \mciteBstWouldAddEndPuncttrue
    \mciteSetBstMidEndSepPunct{\mcitedefaultmidpunct}
    {\mcitedefaultendpunct}{\mcitedefaultseppunct}\relax
    \EndOfBibitem
    \bibitem[Du \latin{et~al.}(2017)Du, Wan, Qu, Cao, Lin, Chen, Lin, and
    Chu]{du2017engineering}
    Du,~H.; Wan,~T.; Qu,~B.; Cao,~F.; Lin,~Q.; Chen,~N.; Lin,~X.; Chu,~D.
    Engineering silver nanowire networks: from transparent electrodes to
    resistive switching devices. \emph{ACS applied materials \& interfaces}
    \textbf{2017}, \emph{9}, 20762--20770, DOI:
    \doi{10.1021/acsami.7b04839}\relax
    \mciteBstWouldAddEndPuncttrue
    \mciteSetBstMidEndSepPunct{\mcitedefaultmidpunct}
    {\mcitedefaultendpunct}{\mcitedefaultseppunct}\relax
    \EndOfBibitem
    \bibitem[Milano \latin{et~al.}(2019)Milano, Porro, Valov, and
    Ricciardi]{milano2019recent}
    Milano,~G.; Porro,~S.; Valov,~I.; Ricciardi,~C. Recent developments and
    perspectives for memristive devices based on metal oxide nanowires.
    \emph{Advanced Electronic Materials} \textbf{2019}, \emph{5}, 1800909, DOI:
    \doi{10.1002/aelm.201800909}\relax
    \mciteBstWouldAddEndPuncttrue
    \mciteSetBstMidEndSepPunct{\mcitedefaultmidpunct}
    {\mcitedefaultendpunct}{\mcitedefaultseppunct}\relax
    \EndOfBibitem
    \bibitem[Coskun \latin{et~al.}(2011)Coskun, Aksoy, and
    Unalan]{coskun2011polyol}
    Coskun,~S.; Aksoy,~B.; Unalan,~H.~E. Polyol synthesis of silver nanowires: an
    extensive parametric study. \emph{Crystal Growth \& Design} \textbf{2011},
    \emph{11}, 4963--4969, DOI: \doi{10.1021/cg200874g}\relax
    \mciteBstWouldAddEndPuncttrue
    \mciteSetBstMidEndSepPunct{\mcitedefaultmidpunct}
    {\mcitedefaultendpunct}{\mcitedefaultseppunct}\relax
    \EndOfBibitem
    \bibitem[Milano \latin{et~al.}(2021)Milano, Pedretti, Montano, Ricci,
    Hashemkhani, Boarino, Ielmini, and Ricciardi]{milano2021materia}
    Milano,~G.; Pedretti,~G.; Montano,~K.; Ricci,~S.; Hashemkhani,~S.; Boarino,~L.;
    Ielmini,~D.; Ricciardi,~C. In materia reservoir computing with a fully
    memristive architecture based on self-organizing nanowire networks.
    \emph{Nature Materials} \textbf{2021}, 1--8, DOI:
    \doi{10.1038/s41563-021-01099-9}\relax
    \mciteBstWouldAddEndPuncttrue
    \mciteSetBstMidEndSepPunct{\mcitedefaultmidpunct}
    {\mcitedefaultendpunct}{\mcitedefaultseppunct}\relax
    \EndOfBibitem
    \bibitem[Rauber \latin{et~al.}(2011)Rauber, Alber, M{\"u}ller, Neumann, Picht,
    Roth, Sch{\"o}kel, Toimil-Molares, and Ensinger]{rauber2011highly}
    Rauber,~M.; Alber,~I.; M{\"u}ller,~S.; Neumann,~R.; Picht,~O.; Roth,~C.;
    Sch{\"o}kel,~A.; Toimil-Molares,~M.~E.; Ensinger,~W. Highly-ordered
    supportless three-dimensional nanowire networks with tunable complexity and
    interwire connectivity for device integration. \emph{Nano letters}
    \textbf{2011}, \emph{11}, 2304--2310, DOI: \doi{10.1021/nl2005516}\relax
    \mciteBstWouldAddEndPuncttrue
    \mciteSetBstMidEndSepPunct{\mcitedefaultmidpunct}
    {\mcitedefaultendpunct}{\mcitedefaultseppunct}\relax
    \EndOfBibitem
    \bibitem[Ferain and Legras(2003)Ferain, and Legras]{ferain2003track}
    Ferain,~E.; Legras,~R. Track-etch templates designed for micro-and
    nanofabrication. \emph{Nuclear Instruments and Methods in Physics Research
        Section B: Beam Interactions with Materials and Atoms} \textbf{2003},
    \emph{208}, 115--122, DOI: \doi{10.1016/S0168-583X(03)00637-2}\relax
    \mciteBstWouldAddEndPuncttrue
    \mciteSetBstMidEndSepPunct{\mcitedefaultmidpunct}
    {\mcitedefaultendpunct}{\mcitedefaultseppunct}\relax
    \EndOfBibitem
    \bibitem[Milano \latin{et~al.}(2020)Milano, Pedretti, Fretto, Boarino,
    Benfenati, Ielmini, Valov, and Ricciardi]{milano2020brain}
    Milano,~G.; Pedretti,~G.; Fretto,~M.; Boarino,~L.; Benfenati,~F.; Ielmini,~D.;
    Valov,~I.; Ricciardi,~C. Brain-inspired structural plasticity through
    reweighting and rewiring in multi-terminal self-organizing memristive
    nanowire networks. \emph{Advanced Intelligent Systems} \textbf{2020},
    \emph{2}, 2000096, DOI: \doi{10.1002/aisy.202000096}\relax
    \mciteBstWouldAddEndPuncttrue
    \mciteSetBstMidEndSepPunct{\mcitedefaultmidpunct}
    {\mcitedefaultendpunct}{\mcitedefaultseppunct}\relax
    \EndOfBibitem
    \bibitem[da~C{\^a}mara Santa Clara~Gomes \latin{et~al.}(2019)da~C{\^a}mara
    Santa Clara~Gomes, Marchal, Abreu~Araujo, and Piraux]{da2019tunable}
    da~C{\^a}mara Santa Clara~Gomes,~T.; Marchal,~N.; Abreu~Araujo,~F.; Piraux,~L.
    Tunable magnetoresistance and thermopower in interconnected NiCr and CoCr
    nanowire networks. \emph{Applied Physics Letters} \textbf{2019}, \emph{115},
    242402, DOI: \doi{10.1063/1.5130718}\relax
    \mciteBstWouldAddEndPuncttrue
    \mciteSetBstMidEndSepPunct{\mcitedefaultmidpunct}
    {\mcitedefaultendpunct}{\mcitedefaultseppunct}\relax
    \EndOfBibitem
    \bibitem[da~C{\^a}mara Santa Clara~Gomes \latin{et~al.}(2019)da~C{\^a}mara
    Santa Clara~Gomes, Abreu~Araujo, and Piraux]{gomes2019making}
    da~C{\^a}mara Santa Clara~Gomes,~T.; Abreu~Araujo,~F.; Piraux,~L. Making
    flexible spin caloritronic devices with interconnected nanowire networks.
    \emph{Science advances} \textbf{2019}, \emph{5}, eaav2782, DOI:
    \doi{10.1126/sciadv.aav2782}\relax
    \mciteBstWouldAddEndPuncttrue
    \mciteSetBstMidEndSepPunct{\mcitedefaultmidpunct}
    {\mcitedefaultendpunct}{\mcitedefaultseppunct}\relax
    \EndOfBibitem
    \bibitem[da~C{\^a}mara Santa Clara~Gomes \latin{et~al.}(2021)da~C{\^a}mara
    Santa Clara~Gomes, Marchal, Abreu~Araujo, Vel{\'a}zquez~Galv{\'a}n, de~la
    Torre~Medina, and Piraux]{da2021magneto}
    da~C{\^a}mara Santa Clara~Gomes,~T.; Marchal,~N.; Abreu~Araujo,~F.;
    Vel{\'a}zquez~Galv{\'a}n,~Y.; de~la Torre~Medina,~J.; Piraux,~L.
    Magneto-Transport in Flexible 3D Networks Made of Interconnected Magnetic
    Nanowires and Nanotubes. \emph{Nanomaterials} \textbf{2021}, \emph{11}, 221,
    DOI: \doi{10.3390/nano11010221}\relax
    \mciteBstWouldAddEndPuncttrue
    \mciteSetBstMidEndSepPunct{\mcitedefaultmidpunct}
    {\mcitedefaultendpunct}{\mcitedefaultseppunct}\relax
    \EndOfBibitem
    \bibitem[Lee \latin{et~al.}(2019)Lee, Gommers, Waselewski, Wohlfahrt, and
    O'Leary]{Lee2019}
    Lee,~G.~R.; Gommers,~R.; Waselewski,~F.; Wohlfahrt,~K.; O'Leary,~A. PyWavelets:
    A Python package for wavelet analysis. \emph{Journal of Open Source Software}
    \textbf{2019}, \emph{4}, 1237, DOI: \doi{10.21105/joss.01237}\relax
    \mciteBstWouldAddEndPuncttrue
    \mciteSetBstMidEndSepPunct{\mcitedefaultmidpunct}
    {\mcitedefaultendpunct}{\mcitedefaultseppunct}\relax
    \EndOfBibitem
    \bibitem[Diaz-Alvarez \latin{et~al.}(2020)Diaz-Alvarez, Higuchi, Li, Shingaya,
    and Nakayama]{diaz2020associative}
    Diaz-Alvarez,~A.; Higuchi,~R.; Li,~Q.; Shingaya,~Y.; Nakayama,~T. Associative
    routing through neuromorphic nanowire networks. \emph{AIP Advances}
    \textbf{2020}, \emph{10}, 025134, DOI: \doi{10.1063/1.5140579}\relax
    \mciteBstWouldAddEndPuncttrue
    \mciteSetBstMidEndSepPunct{\mcitedefaultmidpunct}
    {\mcitedefaultendpunct}{\mcitedefaultseppunct}\relax
    \EndOfBibitem
    \bibitem[Wang \latin{et~al.}(2019)Wang, Wang, Ambrosi, Bricalli, Laudato, Sun,
    Chen, and Ielmini]{wang2019surface}
    Wang,~W.; Wang,~M.; Ambrosi,~E.; Bricalli,~A.; Laudato,~M.; Sun,~Z.; Chen,~X.;
    Ielmini,~D. Surface diffusion-limited lifetime of silver and copper
    nanofilaments in resistive switching devices. \emph{Nature communications}
    \textbf{2019}, \emph{10}, 1--9, DOI: \doi{10.1038/s41467-018-07979-0}\relax
    \mciteBstWouldAddEndPuncttrue
    \mciteSetBstMidEndSepPunct{\mcitedefaultmidpunct}
    {\mcitedefaultendpunct}{\mcitedefaultseppunct}\relax
    \EndOfBibitem
    \bibitem[Araidai and Tsukada(2010)Araidai, and Tsukada]{araidai2010theoretical}
    Araidai,~M.; Tsukada,~M. Theoretical calculations of electron transport in
    molecular junctions: Inflection behavior in Fowler-Nordheim plot and its
    origin. \emph{Physical Review B} \textbf{2010}, \emph{81}, 235114, DOI:
    \doi{10.1103/PhysRevB.81.235114}\relax
    \mciteBstWouldAddEndPuncttrue
    \mciteSetBstMidEndSepPunct{\mcitedefaultmidpunct}
    {\mcitedefaultendpunct}{\mcitedefaultseppunct}\relax
    \EndOfBibitem
    \bibitem[Kuncic \latin{et~al.}(2020)Kuncic, Kavehei, Zhu, Loeffler, Fu,
    Hochstetter, Li, Shine, Diaz-Alvarez, Stieg, \latin{et~al.}
    others]{kuncic2020neuromorphic}
    Kuncic,~Z.; Kavehei,~O.; Zhu,~R.; Loeffler,~A.; Fu,~K.; Hochstetter,~J.;
    Li,~M.; Shine,~J.~M.; Diaz-Alvarez,~A.; Stieg,~A., \latin{et~al.}
    Neuromorphic information processing with nanowire networks. 2020 IEEE
    International Symposium on Circuits and Systems (ISCAS). 2020; pp 1--5, DOI:
    \doi{10.1109/ISCAS45731.2020.9181034}\relax
    \mciteBstWouldAddEndPuncttrue
    \mciteSetBstMidEndSepPunct{\mcitedefaultmidpunct}
    {\mcitedefaultendpunct}{\mcitedefaultseppunct}\relax
    \EndOfBibitem
\end{mcitethebibliography}

\providecommand{\latin}[1]{#1}
\makeatletter
\providecommand{\doi}
{\begingroup\let\do\@makeother\dospecials
    \catcode`\{=1 \catcode`\}=2 \doi@aux}
\providecommand{\doi@aux}[1]{\endgroup\texttt{#1}}
\makeatother
\providecommand*\mcitethebibliography{\thebibliography}
\csname @ifundefined\endcsname{endmcitethebibliography}
{\let\endmcitethebibliography\endthebibliography}{}

\end{document}